# Nonlinear relationships between atmospheric aerosol and its gaseous precursors: Analysis of long-term air quality monitoring data by means of neural networks

Igor B. Konovalov[1]

[1] {Institute of Applied Physics, Russian Academy of Sciences, Nizhniy Novgorod, Russia}

**Abstract**

The nonlinear features of the relationships between concentrations of aerosol and volatile organic compounds (VOC) and oxides of nitrogen (NOx) in urban environments are derived directly from data of long-term routine measurements of NOx, VOC, and total suspended particulate matter (PM). The main idea of the method used for the analysis is creation of special empirical models based on artificial neural networks. These models which are in essence the nonlinear extension of commonly used linear statistical models are believed to provide the best fit for the real (nonlinear) PM-NOx-VOC relationships under different atmospheric conditions. It is believed that such models may be useful in context of various scientific and practical problems concerning atmospheric aerosols. The method is demonstrated by the example of two empirical models created with independent data-sets collected at two air quality monitoring stations at South Coast Air Basin, California. It is shown that in spite of considerable distance between the monitoring stations (more than 50 km) and thus substantially different environmental conditions, the empirical models manifest several common qualitative features. Specifically, it is found that, under definite conditions, the decrease of the level of NOx or VOC may lead to the increase of mass concentration of aerosol. It is argued that these features are caused by the nonlinear dependence of hydroxyl radical on VOC and NOx.

## 1 Introduction

Atmospheric aerosol is an important environmental factor having impact on human health, affecting visibility, and playing a significant role in atmospheric chemistry and climate. The mass concentration of aerosols about the surface is regulated by national air quality standards, implementation of which requires a reliable scientific assessment of possible strategies for its reduction. Such an assessment, however, represents an extremely difficult problem because the evolution of aerosol is driven by complex physical and chemical processes which, in turn, are influenced by a number of environmental factors.

In particular, significant components of typical aerosol in the boundary layer such as sulfate, nitrate, and organic compounds may reach particulate phase as a result of photochemical processes involving certain gaseous species (aerosol precursors) such as $NO_x$, $SO_2$, $NH_3$, and volatile organic compounds (VOC) (e.g., Senfeld and Pandis, 1998). Specifically, the important potential source of the airborne particulate matter (PM) is formation of sulfuric acid in a gas-phase chain mechanism which can be represented by the combined reaction (Stockwell et al., 1990)

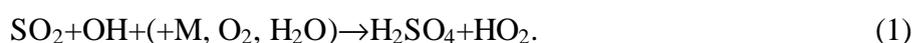
$$SO_2 + OH + (+M, O_2, H_2O) \rightarrow H_2SO_4 + HO_2. \qquad (1)$$

Another potential source of atmospheric PM is the reversible reaction of ammonia with nitric acid

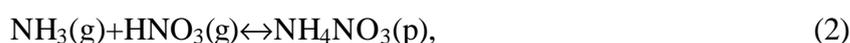
$$NH_3(g) + HNO_3(g) \leftrightarrow NH_4NO_3(p), \qquad (2)$$

where (g) and (p) denote gas and particulate phases, respectively. The $HNO_3$, in turn, is formed in the gas-phase reaction

$$OH + NO_2 \rightarrow HNO_3. \qquad (3)$$

Condensable compounds may be also formed in reactions of certain VOCs with OH, $O_3$, or $NO_3$ (Odum et al., 1996; 1997).

It is obvious from the aforesaid that formation of PM is driven to a significant extent by the hydroxyl radical (OH), the behavior of which is known to be determined by a large number of chemical processes (see, e.g., Stockwell et al., 1997). It is clear also that due to the crucial role of OH in processes of ozone formation, the gas-to-particle conversion processes are closely linked with complex photochemistry of ozone (e.g., Meng et al. 1997).

During two last decades the significant efforts have been devoted to modeling evolution of aerosol from the first principles; that is, the attempts have been made to describe explicitly all important processes affecting formation and removal of air-borne particles (see, e.g., Pilinis and Seinfeld, 1988; Wexler et al., 1994; Binkowski and Shankar, 1995; Jacobson, 1997; Meng et al., 1998; Ackermann et al., 1998; Sun and Wexler, 1998; Nenes et al., 1999). Nonetheless, in spite of the great progress in understanding and representation of aerosol evolution, the comparisons of model results with measurements bear evidence that models still contain significant uncertainties. Moreover, it still remains unknown to which extent the chemical-transport (CT) models are accurate in predictions of responses of aerosol to changes of emissions of potential PM precursors.

Taking into account close chemical coupling between aerosol and ozone, it is useful to note that evaluation of adequacy of CT model predictions of effectiveness of various emission reduction scenarios represents a rather difficult problem in researches of surface ozone as well (e.g., Sillman et al., 1999). In ozone studies, that problem arises, principally, due to the essential non-linearity of dependence of ozone production on concentrations of its precursors (National Research Council (NRC), 1991). In view of this problem, several methods for assessing ozone sensitivity to changes of its precursors have been developed recently (see, e.g., Kleinman et al., 2000) which are based predominantly on the use of data of ambient measurements. On the contrary, similar methods for aerosol that have not yet been proposed. Moreover, though some nonlinear features of PM-$NO_x$-VOC relationships have been discussed (e.g., Meng et al., 1997), it still remains unclear to which extent these features are general.

Thus, at present, there exists the definite need in better understanding of relationships between aerosol and its gaseous precursors, as well as the need in developing additional methods for their identification in the real environment. This paper presents results of the first attempt to study nonlinear features of PM-$NO_x$-VOC relationships directly on the basis of data of ambient measurements. The analysis is conducted by means of the novel original method, the essence of which is creation of empirical nonlinear statistical models based on artificial neural networks. Similar method have been successfully applied recently to studying nonlinear relationships between surface ozone and its precursors (Konovalov, 2002). Though neural networks are widely used in atmospheric researches (see, e.g., the overview by Gardner and Dorling, 1998), they are commonly considered as a "black box" which bears little understandable information of scientific or practical value. However, as it was shown in our previous work (Konovalov, 2002), the possibility of obtaining useful knowledge about physical or chemical properties of a natural system exists if the neural network is trained and analyzed properly.

In this study, neural networks are applied for analysis of data of routine long-term measurements (1980-1995 years) of total suspended particulate (TSP) in two air quality monitoring stations at California. From the point of view of practical usefulness the analysis of fine fraction of aerosol (<2.5 µm) could be of most interest (taking into account US and European standards for $PM_{2.5}$). Unfortunately, at present time the available records of $PM_{2.5}$ measurements accompanied with simultaneous measurements of $NO_x$ and VOC are not long enough for building neural network models aimed at retrieval of PM-$NO_x$-VOC relationships. Nonetheless, we believe that features of these relationships discussed in this paper are fairly robust and applicable to all fractions of PM. It is hoped also that the method suggested in this paper may be successfully applied to fine fraction of aerosol in not very far future.

## 2   Data

This study utilizes data-sets of measurements of TSP, $O_3$, $NO_x$, and total volatile organic compounds (this quantity is referred to below as VOC) made in two air quality monitoring stations at South Coast Air Basin, California during 1980-1995 (California Air Quality Data, 1998). The stations were selected *a priori* based on announced spatial scale of observations (only stations with the urban scale (4-50 km) of ozone and $NO_x$ observations were considered) and availability of synchronous measurements of all components noted above. The urban scale means that both distribution of sources of pollutants and topographical situation in the neighborhoods of the station are such that the appearance of strong spatial gradients of the corresponding pollutants is unlikely. The only stations satisfying to these criteria turned out to be Riverside-Rubidoux ($33^0 58'55''$N, $117^0 22'21''$W), Azusa ($34^0 08'10''$N, $117^0 55'26''$W) and Reseda ($34^0 11'57''$N, $118^0 31'58''$W). However, since a series of TSP measurements at Reseda is significantly shorter than those at the other two stations (at Reseda, the measurements were conducted only in 1980-1986), the data-set from this station was not analyzed. Note that data series used in this analysis ended in 1995, as the measurements of VOC at South Coast Air Basin were cancelled that year. The measurements of TSP (they present daily values of mass concentration) were conducted rather irregularly. The most frequent interval of subsequent measurements were 6 days; however, there is a number of much more extended gaps in the data series. The measurements of gaseous species were much more regular and were conducted usually each hour. Nonetheless, corresponding data series also contain a number of gaps. The hourly data for gaseous species were averaged from 10 to 18 o'clock each day of measurements. Such averaging were performed in order to get unique value of each component per a day and to suppress effects of possible random variations of species concentrations within the course of the day. Note also that since the aim of this study is to analyze manifestations of photochemical processes which are most active during light period of day, it is natural to consider dependence of PM on the day-time rather than night-time or daily mean concentrations of gaseous pollutants. No doubt, it would be advantageous to use TSP data for the same period of a day as $NO_x$ and VOC data, but they are not available. Nonetheless, it is reasonable to hope that neural networks themselves are able to retrieve a part of variations of daily mean TSP caused by variations of day-time concentrations of $NO_x$ and VOC.

The air quality data were matched by daily mean meteorological data from Climate Prediction Center, National Ocean and Atmospheric Administration (http://www.cpc.ncep.noaa.gov/). Again, it is reasonable to assume that the use of the meteorological data averaged over the same period of day as gaseous species data would facilitate to the better quality of the empirical models. Unfortunately, free hourly meteorological data were not found.

Finally, the data-sets analyzed in this study include 1880 and 1032 days of measurements at Azusa and Riverside-Rubidoux stations, respectively. The data have not been pre-selected in any way.

## 3  Methodology

The basic assumption in creation of the empirical models for aerosol is that mass concentration of TSP (this quantity is referred to below simply as PM) can be approximately represented as a sum of a nonlinear function (*f*) of *n* other measured values $x_i$, $i=1,\ldots,n$ and some random value (***d***), variations of which cannot be accounted for by variations of the arguments of *f*:

$$PM = f(x_1,\ldots,x_n) + \boldsymbol{d}, \qquad (4)$$

The proper choice of the arguments $x_i$ is the most critical point in providing usefulness and reliability of the empirical models. Obviously, the arguments should include, in particular, such values, the dependence on which is to be studied. In this study, such values are gaseous precursors of PM, that is, $NO_x$ and VOC. It would be of interest to consider other important gaseous precursors such as $NH_3$ and $SO_2$, but adequate contemporary measurements of these species are not available. It is prudent also to take into account other important factors which, presumably, can significantly influence studied relationships. However, a number of such factors should not be too large. Indeed, it is obvious that larger number of arguments of a function, more data are needed to approximate it with the same accuracy. Taking into account the aforesaid, we have included in the PM empirical models only four meteorological values: temperature, specific humidity, and two horizontal components of wind speed. The empirical models with an additional argument, for example, total cloud cover which was included in our ozone empirical models (Konovalov, 2002), were found to demonstrate higher diversity between them in properties discussed in Section 4 than the models without this argument, what is indicative of the deficit of data. Thus we believe that the models with seven arguments are less adequate and thus such models will not be considered here. Nonetheless, it was found that empirical models with seven arguments retain the most important features of PM-$NO_x$-VOC relationships retrieved from the models with six arguments.

It is assumed further that *f* can be approximated by means of a neural network of the perceptron type with one hidden layer (see, e.g., Bishop, 1995). That the perceptron with just one hidden layer (given large enough number of neurons) is capable of approximating any measurable function to any desired degree of accuracy has been proven by Hornik et al. (1989). Mathematically, the perceptron $p(\mathbf{X})$ with one hidden layer can be represented as follows:

$$p(\mathbf{X}) = \sum_{j}^{N_N}\left(w_j g_j(\mathbf{X}) + w_0\right), \ g_j(\mathbf{X}) = \frac{1}{1 + exp[-\sum_{i=1}^{n}(\overline{w_{ij}} x_i + \overline{w_{0j}})]}, \qquad (5)$$

where $\mathbf{X}=(x_1,\ldots,x_n)$ is a vector of input values (that is, a vector of arguments of the function f), $w_j$, $w_0$, $\overline{w_{ij}}$, $\overline{w_{0j}}$ are weight coefficients, $N_N$ is a number of neurons in the network.

A procedure by which appropriate values of weights are determined is one of the key elements of the methodology of this study. Such a procedure is usually referred to as a training algorithm. Here a standard backpropagation algorithm (e.g., Reed and Marks, 1999) is used in combination with the early stopping method (Nelson and Illingworth, 1991; Sarle, 1995). The early stopping is usually used to provide such an important property of a neural network as generalization. For most applications generalization means that outputs of the network are able to approximate target values with given inputs that are not in the training set (Sarle, 2002). In case of this study, generalization means that a neural net learns the regular relationships (*f*) which are valid for any arbitrary chosen subset of the considered data-set and disregards noisy variations ($\delta$). Implementation of early stopping in this study includes the following steps. First, the sequence of days in the data-sets was randomly mixed. Thereafter each of the data-sets were divided into two equal subsets: training and validation ones. The net is trained with the training subset but the error of prediction which is defined as

$$e = \left[ \frac{2}{N_d} \sum_{i=1}^{Nd/2} \left( PM^{ob} - PM^n \right)^2 \right]^{1/2}, \quad (6)$$

where $PM^{ob}$ and $PM^n = p(\mathbf{X})$ are observed and predicted PM levels, and $N_d$ is number of training cases (that is, the total number of days presented in the data-set), is controlled with the validation subset. The iterations are stopped as $\varepsilon$ reaches minimum.

Direct application of this technique for creation of empirical models in the form of a single perceptron (5) meets two main problems. First, it cannot be neglected a priori that perceptrons with different numbers of neurons ($N_N$) may give qualitatively different results. Second, since the training of neural networks is an iteration process, the results may depend on initial assumption regarding values of weights. To resolve these problems, we considered first, to what degree changes of $N_N$ may affect quality of approximations and important features of the relationships analyzed. It was found, in particular, that the dependence of $\varepsilon$ (see (6)) on $N_N$, averaged over large number of single perceptrons with different random initial weights, is saturating such that systematic differences between perceptrons with different $N_N$ greater, at least, than 20 are quite negligible. The similar result were obtained for dependencies of derivatives of PM with respect to $NO_x$ and VOC, evaluated from (5), on $N_N$. Values and, especially, signs of these derivatives are the most important characteristics in the context of this study. Based on these results, it was concluded that the dependence of the approximations on $N_N$ is not of any significance if $N_N$ is greater, at least, than 20. In order to minimize any hypothetical uncertainties associated with choice of definite value of $N_N$ to maximum degree, it was decided to include in the models perceptrons with different $N_N$ ranging from 21 to 30.

To get rid of uncertainty associated with initial values of weights, a large enough set (50) of perceptrons with random initial weights were trained for each value of $N_N$ and a few (5) "best" (having minimum $\varepsilon$) networks selected from that set were included in the empirical model. Finally, the desired empirical model is represented as the mean of outputs of the set of 50 networks, which includes five "best" networks for each $N_N$ from 21 to 30:

$$PM = \frac{1}{50} \sum_{N_N=21}^{30} \sum_{i=i_{b1}}^{i_{b5}} p(\mathbf{X}, N_N, i), \quad (7)$$

where $i_b$ are the indexes of the best nets.

That the uncertainties in results of the empirical models, associated with the choice of values $N_N$ and initial values of weights are indeed quite small is illustrated in Fig. 1. This figure presents comparisons of evaluations of first derivatives of PM with respect to VOC obtained from pairs of empirical models created in general accordance to (7) but with $N_N$ ranging from 21 to 25 and from 26 to 30. Each point in the figures represents one day in the training sub-set of the data-sets. It is seen that each pair of independently created versions of the empirical models gives very close results. In particular, fractions of days for which estimated signs of $\partial PM/\partial VOC$ values are different in the different versions are only 2.1% and 3.1% for the Azusa and Riverside models, respectively. It is reasonable to expect that the uncertainty of the full empirical models (7), associated with the discussed factors, should be even smaller than that of their component parts.

The empirical model performances in representation of variability of PM are demonstrated in Fig. 2, which show measured values of PM (from the training sub-sets) versus those evaluated by the models. It is quite natural that the models do not reproduce the measured values exactly since they are intended to catch only a part of total variability, associated with changes of the factors represented by the model arguments. It is seen, however, that this part is quite substantial. This fact bear evidence that the model arguments indeed include values which strongly influence the PM level. The results for the validation subsets are very insignificantly worse.

The main idea of the analysis of the empirical models consists in the following. It is assumed that empirical models provide as precise as possible fits of actual PM-$NO_x$-VOC relationships under various conditions determined by values of the arguments. The qualitative features of these relationships under the given conditions may be learned via evaluation of various derivatives of PM with respect to VOC and $NO_x$. For example, positive first derivative PM with respect to VOC means that the increase of the level of VOC, most probably, would cause the increase of the level of PM under the given conditions; that is, the PM level demonstrates positive sensitivity to changes of the VOC level. It should be noted, however, that positive first derivatives of PM with respect to VOC or $NO_x$ may reflect simply the natural correlation between the levels of *primary* PM and gaseous pollutants as they mainly originate from the same emission sources. The positive second derivative of PM with respect to VOC under the same conditions means that the sensitivity would increase with the increase of VOC. The derivatives that present interest may be evaluated, in principle, under any values of argument from the range of observed conditions. However, the most accurate evaluations are expected for conditions observed at days presented in the training sub-set of the data-sets. The special investigation has shown that the evaluations of individual nets forming the empirical models for the conditions that are not in the training subset, including those that are presented in the validation sub-sets, are much more dispersed. This is, presumably, due to the fact that each individual neural network included in the model is learned for the conditions that are in the training sub-set but has more freedom under other conditions. Thus it was decided to analyze only the evaluations made for the days presented in the training sub-sets.

It is rather obvious that PM-$NO_x$-VOC relationships in the real environment under windy conditions may be affected by peculiarities of distributions of sources of pollutants. Thus, after creation of the models, it was found to be useful to make the approximation of no wind conditions. That is, all model evaluations were made with values of wind speed in (7) to be zero. Such approximation is possible as similar values of pollutant concentrations, temperature and humidity can be associated with different magnitudes and directions of wind. No wind conditions allow us to identify the effects associated with local processes of formation of secondary PM.

The obtained evaluations are analyzed subsequently in order to find some associations between values (and signs) of different derivatives and argument values. For example, it is considered, whether days with negative values of first derivatives of PM with respect to VOC are more associated with lower temperatures than days with the positive derivatives, or with higher temperatures. If we had only one empirical model, any association of such type would be of little utility. Indeed, it would not be possible to conclude then, whether the observed association is the indication of some real regularity, or it is just a result of poor quality of the model. However, when the same result is found with other model created using the independent data-set collected in a different environment, we have much more grounds to believe that this result is indeed a consequence of some general regularity. It should be taken into account that untrained or trained with the random data-sets neural networks give random outputs and thus cannot demonstrates any significant correlation either between different output values, or between output values and values of arguments.

# 4 Results

## 4.1 PM-$NO_x$ relationship

The rate of gas-to-particle conversion due to the reactions (1)-(3) is proportional to concentration of hydroxyl radical. Thus it can be expected that the increase of the OH level under other constant conditions would lead to the increase of PM production, and vice versa. It is well known that OH itself depends on $NO_x$ in the essentially non-linear manner (see, for example, Kley et al., 1999). Specifically, OH increases with the increase of $NO_x$ under low enough $NO_x$ level, but may decrease under higher $NO_x$ level. Thus, there is a possibility that the decrease of $NO_x$ may actually lead to the increase of the level of PM. Though this idea is rather obvious, any experimental demonstration of this possibility, as far as we know, has not been discussed in peer reviewed literature earlier.

In order to check this idea, we have evaluated (using (7)) first derivatives of PM with respect to $NO_x$ ($\partial PM/\partial NO_x$) for each day presented in the training sub-sets of the data-sets. It was found that though majority of those days are associated with positive $\partial PM/\partial NO_x$, there are a few of them (8 days from 940 ones in the training sub-set for the Azusa and 6 days from 516 ones for Riverside) associated with $\partial PM/\partial NO_x<0$. In order to understand the origin of negative $\partial PM/\partial NO_x$ in the empirical models, some additional analysis was undertaken.

First, we considered dependencies of PM on $NO_x$ for all days with $\partial PM/\partial NO_x<0$. Such dependencies are shown in Fig. 3. They were calculated from (7) assuming that all model arguments except $NO_x$ have fixed values corresponding to the days considered. The general features of all these dependencies are the presence of growing parts under lower levels of $NO_x$ and falling parts under higher levels of $NO_x$. The same features are characteristic for the dependence of OH on $NO_x$ (Kley et al., 1999). Some of the dependencies for the Azusa station possess growing parts also under highest values of $NO_x$. Most likely, this growth is an artifact of the empirical model and caused by insufficient number of data with high values of $NO_x$. Indeed, the days with $NO_x>120$ ppb constitute less than 3% of total number of days in the data-set for the Azusa station.

Second, it was found that at both stations, days with $\partial PM/\partial NO_x<0$ are associated, on the average, with significantly higher concentrations of $NO_x$ (107 ppb at the Azusa station and 110 ppb at the Riverside station), higher concentrations of VOC (2.43 ppmC and 2.61 ppmC) and lower temperatures (18.8 $^0$C and 14.6$^0$C), as compared with corresponding values averaged over the whole data sets (58.5 ppb and 31.6 ppb; 2.38 ppmC and 2.21 ppmC; 22.2$^0$C and 20.4$^0$C at the Azusa and Riverside stations, respectively). Regarding influence of VOC level on values of $\partial PM/\partial NO_x$ it was found additionally that at both stations, all days except one at the Azusa station are associated with positive values of mixed second derivative $\partial^2 PM/\partial VOC\partial NO_x$. It means that the increase of the VOC level actually facilitates to the less negative values $\partial PM/\partial NO_x$. Thus the noted above association of days when $\partial PM/\partial NO_x<0$ with higher concentrations of VOC is, most probably, just a result of positive correlation between the ambient VOC and $NO_x$ levels. Any common, for both models, association between the signs of $\partial PM/\partial NO_x$ and second order derivatives of PM with respect to temperature and $NO_x$ were not found. This fact, most probably, indicates significant uncertainty of evaluations of these derivatives. It is known from the theory (see, for example, Kleinman, 1994; Sillman, 1999) that higher $NO_x$ level, lower insolation (which can be expected to be associated with lower temperature) and lower VOC level facilitate to the $NO_x$-rich regime, when radical level falls with the increase of $NO_x$. Thus, the noted results of the empirical models are, at least, do not contradict to the hypothesis that negative sensitivity of PM to $NO_x$ has the photochemical origin.

Finally, let us discuss the reasons of smallness of a fraction of days with $\partial PM/\partial NO_x<0$. We see two possible reasons. On the one hand, as $NO_x$ itself can be incorporated into particulate material according to the mechanism (2), (3), it is obvious that the ridge in dependence of PM on $NO_x$ should be shifted toward higher values of $NO_x$, in comparison with the ridge in dependence of OH on PM under the same conditions, and thus a fraction of days with $\partial PM/\partial NO_x<0$ should be smaller than a fraction of days with $\partial OH/\partial NO_x<0$. On the other hand, in the real environment, the level of $NO_x$ may correlate with intensity of the primary emission of PM and/or with the levels of other precursors of PM, such as $NH_3$ or $SO_2$, which are not included in the empirical models. Obviously, such correlation would lead to a persistent positive shift in $\partial PM/\partial NO_x$ values evaluated by the empirical models. Nonetheless, in our opinion, even with regard of quantitative uncertainties of the empirical model evaluations, the possibility of negative sensitivity of PM level to change of $NO_x$ is established here rather firmly.

## 4.2 PM-VOC relationship

### 4.2.1 Empirical results

The first evidence for the essential non-linearity of PM-VOC relationship was found by Meng et al. (1997) in results of their PM-CT model simulation of changes of $PM_{2.5}$ level in response to changes in emission rates of $NO_x$ and VOC at Riverside, California, under conditions of 28 August 1987. They found, in particular, that the decrease of VOC emission rate may lead either to the increase or decrease of $PM_{2.5}$ level, depending on the base case for $NO_x$ and VOC emission rates. Specifically, greater (by absolute value) negative sensitivity of PM level to changes of VOC level was found, in most cases considered, under higher levels of VOC. Unfortunately, fruitful comparison of the theoretical results by Meng et al. (1997) with our empirical results is seriously hindered due to the following reasons. (1) The model by Meng et al. and our empirical models use different measures of PM ($PM_{2.5}$ and TSP, respectively). (2) The arguments of the empirical models are surface concentrations of $NO_x$ and VOC. Thus the empirical models cannot account, for example, for the possible changes of $NO_x$ surface concentration after changes of VOC emission rate due to chemical interaction between VOC and $NO_x$. (3) The results of Meng et al. concern maximum 1-hour average concentration of PM, whereas our empirical models deal with daily averaged level of PM. Nonetheless, the consideration of dependence of PM on VOC given by the empirical model for the Riverside station under conditions of August 28, 1987 (not presented here) reveals that this dependence is also essentially non-monotonous and that high enough values of VOC correspond to negative sensitivity of PM to VOC, whereas the lower ones facilitate to the positive sensitivity.

Empirical model evaluations of the first derivative of PM with respect to VOC ($\partial PM/\partial VOC$) shows that some of the days presented in the data-sets are associated with negative sensitivity of PM to changes of VOC and others with positive one. These evaluations are plotted in Fig. 4 and 5 against temperature and concentration of VOC observed at corresponding days, respectively. It is seen, in particular, that noticeable fractions of the days (namely, 6.7% at the Azusa station and 31.4% at the Riverside station) have negative values of $\partial PM/\partial VOC$. Furthermore, the points with $\partial PM/\partial VOC<0$ in Fig. 4 are not distributed uniformly but tend "to avoid" definite ranges of moderately warm temperatures, thus forming a "butterfly-like" structures. Solid curves represent polynomial (of the $4^{th}$ order) fits of the data and illustrate non-linear character of dependence (in the statistical sense) of $\partial PM/\partial VOC$ on temperature. These features allow us to hypothesize that different processes drive behavior of PM under low and high temperatures. This hypothesis finds further confirmation in Fig. 5, where points are separated into high-temperature and low-temperature "modes": the threshold temperatures are chosen to be equal to $18^0C$ at the Azusa station and $22^0Ñ$ at the Riverside station and marked in Fig. 4 by dash-dot vertical lines. Specifically, it is seen that, under low enough VOC level, the high-temperature mode is associated predominantly with positive $\partial PM/\partial VOC$, whereas the majority of the points with negative ∂PM/∂VOC belong to the low-temperature mode. In contrast, under high enough VOC level, the most part of the points with ∂PM/∂VOC<0 belong to the high-temperature mode. The noted features give an idea to perform separate analysis of PM-VOC relationships for the different temperature modes.

Table 1 presents average observed values of PM, VOC, $NO_x$, T, and RH, calculated separately for high temperature and low temperature modes, separated additionally with regard to the sign of predicted value of ∂PM/∂VOC in the respective days. This table presents also day fractions (of total number of days in the given temperature mode) satisfying to certain conditions concerning values of temperature, the sign of ∂PM/∂VOC, and the sign of corresponding second order derivatives of PM. It is worth to note the following general for both models features. First, in the low (high) temperature mode, the average values of PM, VOC, $NO_x$, and T is less (greater) in days with negative ∂PM/∂VOC than in the days with positive ∂PM/∂VOC. Second, the minority (majority) of days have negative values of $∂^2PM/∂VOC^2$ and $∂^2PM/∂T∂VOC$ in the low (high) temperature mode. Values of relative humidity do not demonstrate general for both models changes with changes of the sign of ∂PM/∂VOC. There may be only two reasons for that. First, at least one of the models did not "learn" the relations between ∂PM/∂VOC and RH satisfactorily. Second, these relations are indeed qualitatively different in different environments. At the given state of knowledge of these relationships, it is not possible to determine which of two reasons is true. The similar situation takes place with the behavior of $∂^2PM/∂NO_x∂VOC$. The only common feature is that the ratio of number of days having both negative $∂^2PM/∂NO_x∂VOC$ and ∂PM/∂VOC to number of days with negative ∂PM/∂VOC (this ratio can be obtained by dividing the number in the 8$^{th}$ row of the table to that in the 1$^{st}$ row) is significantly less than 0.5 at the high temperature mode. An attempt to identify physical and chemical reasons for noted features is described in the following section.

### 4.2.2  Discussion of PM-VOC relationship

First of all, it is necessary to note that the following discussion does not pursue the notion of exact identification of the processes which cause the noted above features of PM-VOC relationships. Such identification is out of the scope of this paper and would require in-detail consideration of complex processes that drive evolution of PM at the given site. Here, we would like to formulate and substantiate, wherever possible, the most plausible hypotheses for the explanation of the empirical relationships.

We believe that the main factor driving formation of secondary PM in the high-temperature mode is the behavior of OH radical. The behavior of OH radical is directly related to the behavior of ozone and underlines qualitative nonlinear features of $O_3$-$NO_x$-VOC relationships which were extensively studied during two last decades (see, for example, NRC, 1991). Correspondingly, the key processes responsible for the essential general nonlinear features of the dependence of OH on VOC and $NO_x$ have also been identified earlier. In particular, those features were interpreted by Sillman (1990, 1995) by means of the mechanistic model based on the equation of the radical budget. It is demonstrated below that the main empirical features of PM-VOC relationships discussed above are closely associated with features of OH-VOC relationships given by the Sillman's model. Note that the use of the mechanistic model in this study appears to be quite natural, as our desire here is to appeal to the most general qualitative properties of the tropospheric photochemistry. For the sake of clarity of the following discussion, it is worth to give here brief description of that model, with inessential modifications.

It includes the schematic representation of the sequence of the essential processes determining the sources and sinks of the radicals OH, HO$_2$ and RO$_2$. (The latter denotes various organic radicals.) Specifically, RO$_2$ is formed in the reaction of a hydrocarbon (RH) with OH,

(R1) $\qquad\qquad\qquad$ RH+OH→RO$_2$+H$_2$O

The instantaneous equilibrium of RO$_2$ radicals is provided by the reaction

(R2) $\qquad\qquad\qquad$ RO$_2$+NO→HO$_2$+NO$_2$+products,

which also provides the source of HO$_2$ radical, whose concentration is balanced by reaction

(R3) $\qquad\qquad\qquad$ HO$_2$+NO→OH+NO$_2$

A magnitude of the sum of concentrations of OH, HO$_2$ and RO$_2$ radicals (this sum is called as odd hydrogen) is determined by the equation of the radical budget,

$$S_H = (R4)[NO_2][OH] + \left( (R5)\left(\frac{(R1)[RH]}{(R3)[NO]}\right)^2 + \frac{(R6)(R1)^2[RH]^2}{(R3)(R2)[NO]^2} \right)[OH]^2, \qquad (8)$$

where it is assumed that the source of odd hydrogen, S$_H$, is balanced by its sinks provided by the reaction of OH with NO$_2$ (see (3)) with the rate coefficient (R4), and the radical-radical reactions,

(R5) $\qquad\qquad\qquad$ HO$_2$+HO$_2$→H$_2$O$_2$+O$_2$

(R6) $\qquad\qquad\qquad$ RO$_2$+HO$_2$→ROOH+O$_2$.

In Eq. (8), square brackets denote concentrations of species. S$_H$ may be further parameterized as

$$S_H = S_1 + S_2[RH], \qquad (9)$$

where it is assumed that the main processes responsible for the radical production are photolysis of ozone and aldehydes.

Assuming that the dependence of ratios of NO$_2$ and NO to NO$_x$ on VOC and NO$_x$ can be neglected, concentration of OH can be obtained by resolving the quadratic equation. With a little manipulation Eq. (8) can be represented in the most general, non-dimensional form as

$$1 + \hat{a}x = a y v + \frac{x^2}{y^2} v^2, \qquad (10)$$

where

$$v = z[OH], \quad z = \frac{[RH]_0 (R1)}{[NO]_0 S_1^{1/2}} \left( \frac{(R5)}{(R3)^2} + \frac{(R6)}{(R3)(R2)} \right)^{1/2},$$

$$x = [RH]/[RH]_0, \quad y = [NO_x]/[NO_x]_0, \qquad (11)$$

$$a = \frac{(R4)[NO_2]_0}{zS_1}, \quad \beta = S_2[RH]_0/S_1,$$

and subscript "$_0$" denotes some typical values of the species concentrations.

In typical polluted environments, the sources of radicals due to photolysis of ozone and aldehydes are of the same order of magnitude, so typical value of $\beta$ is about of unity. Thus the dependence of OH on VOC is almost exclusively determined by the parameter $\alpha$.

Specifically, when $\alpha \gg 1$, OH increases with the increase of VOC ($\partial \nu/\partial x \approx \beta/\alpha y > 0$). However, as $\alpha$ decreases, $\partial OH/\partial VOC$ can become negative. For example, when $\alpha \ll 1$, $\nu \approx y(1+\beta x)^{1/2}/x$ and $\partial \nu/\partial x < 0$. It is evident from (11) that less value of $\alpha$ corresponds to situations with greater overall reactivity of hydrocarbons (i.e., greater (R1)) and greater insolation and ozone concentration (i.e., greater $S_1$)). It is clear that in the real atmosphere, those are associated with greater, on average, temperature. These features may be responsible for the appearance of the negative sensitivity of PM to changes of VOC at the high-temperature mode, the association of days with $\partial PM/\partial VOC$ with higher temperatures and the dominance of negative values of $\partial^2 PM/\partial T \partial VOC$ at those days.

It is obvious also from (10) that a similar "switch" from $\partial \nu/\partial x > 0$ to $\partial \nu/\partial x < 0$ under fixed parameter values can be caused by either an increase of x, or a decrease of y. This feature may be responsible for the empirical facts that the high-temperature mode is mostly associated with negative values of $\partial^2 PM/\partial VOC^2$, and days with $\partial PM/\partial VOC<0$ at the high-temperature mode are associated with higher values of VOC and, mostly, with positive values of $\partial^2 PM/\partial NO_x \partial VOC$. Note that x and y are involved in (10) asymmetrically. Specifically, in the situation with $\alpha \gg 1$, the increase of y leads to the decrease of $\partial \nu/\partial x$, that is $\partial^2 \nu/\partial x \partial y < 0$, in contrast to the situation with $\alpha$ about unity, when $\partial^2 \nu/\partial x \partial y > 0$. At the same time, it is can be seen that $\partial^2 \nu/\partial x^2$ is negative in both these situations and becomes positive only when $\alpha \ll 1$. This observation, with the additional assumption that in the real situations $\alpha$ is greater or insignificantly less than unity, may account for the empirical fact that domination of $\partial^2 PM/\partial VOC^2 < 0$ takes place independently on the sign of $\partial PM/\partial VOC$ at the high temperature mode, whereas domination of $\partial^2 PM/\partial NO_x \partial VOC > 0$ is found only in days with $\partial PM/\partial VOC < 0$.

We believe that conditions associated with the low-temperature mode correspond, generally, to $\alpha \gg 1$. Indeed, the CT model evaluations (see, for example, Milford et al., 1994) show that significant areas at South Coast Air Basin (SCAB) are under $NO_x$-saturated conditions (that is, in terms of our model, with $\alpha > 0$) even in the hot and sunny days. Thus, it is reasonable to expect that significant decrease of insolation in colder seasons would drive the photochemistry of SCAB into conditions with $\alpha \gg 0$. Under these conditions concentration of OH is relatively small and, as a result, the behavior of PM in the low-temperature mode is affected by non-chemical processes to much greater extent than that in the high-temperature mode. In particular, it can be speculated that dominating positive values of $\partial^2 PM/\partial VOC^2$ in the low-temperature mode are associated, in part, with a physically evident possibility that the increase of total concentration of hydrocarbons is associated with the increase of number of individual semi-volatile hydrocarbons reaching the saturation. Consequently, the increase of measured total concentration of volatile hydrocarbons (that is, VOC) may be associated with the increase of the ratio of total concentration of hydrocarbons in the condensed phase to concentration of VOC, and as a result, with positive values of $\partial^2 PM/\partial VOC^2$. However, the same effect can be caused by the chemical processes of formation of PM via oxidation of volatile hydrocarbons [Odum et al., 1996; 1997]. The rate of formation of PM due to these processes is proportional to the product of concentrations of VOC and OH, that is, in terms of Eq. (8), it is proportional to $x\nu$. It is easy to see that in situation with $\alpha \gg 1$, $\partial^2 x\nu/\partial x^2 > 0$. This process may account also for substantial fraction of days with $\partial^2 PM/\partial VOC^2 > 0$ at the high-temperature mode.

The most speculative point of this discussion concerns the reasons for negative sensitivity of PM to VOC under very low, relatively, temperatures. We failed to find any chemical reason for this, consistent with majority of the empirical facts. It can be assumed that the increase of concentration of VOC, associated with the increase of concentration of condensed semi-volatile hydrocarbons, hinders condensation of inorganic semi-volatile species such as $NH_4NH_3$ by providing a hydrophobic film on the surface of aerosol particles and thus leads, effectively, to the increase of saturation vapor pressure, or to the decrease of value of the Henry's law constants. At present, we do not aware of experimental or theoretical data that would allow us to estimate actual possibility and significance of this effect. Under higher temperatures, which are associated, presumably, with less $\alpha$ and, consequently, larger values of $\partial OH/\partial VOC$ ($\partial v/\partial x \approx \beta/\alpha y$ when $\alpha \gg 1$), this effect is suppressed, probably, by the more rapid increase of concentration of semi-volatile inorganic species with the increase of VOC. This effect may be suppressed also because of the increase of fraction of organic aerosol with the increase of concentration of VOC under the same temperature. Anyway, it is evident that results of the empirical models concerning the low-temperature conditions pose very interesting questions for further researches.

## 5  Summary and conclusions

This paper presents results of the first attempt of analysis of nonlinear relationships between airborne particulate matter and its gaseous precursors based on observational data. The analysis of the long term monitoring data at South Coast Air Basin, California, is performed by means of original method which assumes creation of the empirical models based on the artificial neural networks. Such models does not involve any a priori assumptions regarding character of analyzed relationships. The reliability of results of this study is confirmed by the fact that two empirical models created using independent data-sets which are obtained in two different monitoring stations separated by a distance more than 50 km manifest a number of common qualitative features of the relationships between PM, VOC and $NO_x$. The models are assumed to provide the best fit for the real PM-VOC-$NO_x$ relationships under different observed values of temperature, relative humidity, and horizontal wind speed. In order to suppress effects caused by transport of PM from different areas and to emphasize the features of PM-VOC-$NO_x$ relationships determined by the local formation of PM, the approximation of zero wind speed is enforced in the models. With this approximation, the models are used to evaluate different derivatives of PM with respect to VOC and $NO_x$ under conditions corresponding to a number of days presented in the data-sets. These evaluations are used further to identify general, in the statistical sense, regularities of PM-VOC-$NO_x$ relationships.

It is found, in particular, that there are conditions when PM demonstrates a negative sensitivity to changes of $NO_x$, or VOC (that is, when first derivatives of PM with respect to $NO_x$ or VOC are negative). It is found also that negative sensitivity of PM to $NO_x$ is facilitated by higher ambient $NO_x$ concentrations and lower temperatures and VOC concentrations. These features are consistent with the hypothesis that the non-linearity of PM-$NO_x$ relationships is caused by the corresponding non-linearity of the dependence of concentration of hydroxyl radical (OH) on $NO_x$.

The qualitative features of relationships between PM and VOC are found to be dependent on temperature. In view of this dependence it is turned out to be fruitful to separate analyzed days into two "modes" with respect to value of temperature (T) observed at those days: the high-temperature mode and the low-temperature one, with the threshold temperature to be about $20^0C$. Consequently, the PM-VOC relationships are analyzed for each temperature mode separately. It is found that in the low-temperature mode, the appearance of negative sensitivity of PM with respect to VOC is facilitated by lower temperatures and VOC concentrations, whereas positive and greater values of the sensitivity correspond to higher temperatures and VOC concentrations. In the high-temperature mode, the overall picture is inverse: the appearance of negative sensitivity of PM with respect to VOC is facilitated by higher temperatures and VOC concentrations, whereas positive and greater values of the sensitivity correspond to lower temperatures and VOC concentrations.

It is argued that the most significant features of PM-VOC-$NO_x$ relationships in the high-temperature mode are closely associated with the similar features of OH-VOC-$NO_x$ relationships, whereas the behavior of aerosol in the low temperature mode is strongly perturbed by peculiarities of the processes of condensation of primary and secondary semi-volatile compounds. In our opinion, the empirical results of this study provide the rich "food" for the further theoretical analysis.

The main condition of future successful applications of the methodology discussed in this paper is availability of long enough concurrent measurements of PM and its main gaseous precursors. Up to now, there were only a few stations satisfying to this criteria. Moreover, unfortunately, the necessary measurements of VOC at air pollution monitoring stations at California were almost totally cancelled in 1995 and thus it is not possible to apply this methodology for assessment PM-$NO_x$-VOC relationships at the present-day situation. Nonetheless, in the near future, the same methodology may be applied to the analysis of data of $PM_{2.5}$ and PM precursors measurements at the Photochemical Assessment Monitoring Stations (PAMS) at US (US Environmental Protection Agency, 1994), which have begun to operate in recent years. Anyway, the results of this study emphasize the importance and fruitfulness of concurrent and long-term measurements of coupled parameters of the atmospheric environment.

The results of this study indicate that the elaboration of the effective strategy for reduction of the PM level can present a rather difficult problem in view of essential non-linearities of relationships between PM and its gaseous precursors. These non-linearities should be studied further.

**Acknowledgements**
This work was supported in part by the Russian Foundation for Basic Researches, grants 99-02-16162 and 02-02-17080.

Table 1. Average characteristics of days with ∂PM/∂VOC<0 and ∂PM/∂VOC≥0 for different temperature modes

| | Azusa | | | | Riverside-Rubidoux | | | |
|---|---|---|---|---|---|---|---|---|
| | T<18$^0$C | | T≥18$^0$C | | T<22$^0$C | | T≥22$^0$C | |
| | $\frac{\partial PM}{\partial VOC}<0$ | $\frac{\partial PM}{\partial VOC}\geq 0$ | $\frac{\partial PM}{\partial VOC}<0$ | $\frac{\partial PM}{\partial VOC}\geq 0$ | $\frac{\partial PM}{\partial VOC}<0$ | $\frac{\partial PM}{\partial VOC}\geq 0$ | $\frac{\partial PM}{\partial VOC}<0$ | $\frac{\partial PM}{\partial VOC}\geq 0$ |
| Fraction of days (%) | 17.2 | 82.8 | 4.0 | 96.0 | 41.9 | 58.1 | 17.0 | 83.0 |
| PM (μg/m$^3$) | 68.7 | 79.8 | 138.5 | 133.0 | 72.2 | 130.6 | 196.9 | 177.4 |
| VOC (ppmC) | 1.96 | 2.28 | 3.73 | 2.38 | 1.96 | 2.31 | 2.84 | 2.14 |
| NO$_x$ (ppb) | 42.9 | 61.6 | 69.2 | 58.0 | 27.1 | 39.8 | 37.0 | 26.1 |
| T ($^0$C) | 9.8 | 13.6 | 28.1 | 25.3 | 12.6 | 16.9 | 28.2 | 26.7 |
| RH (%) | 52.7 | 68.8 | 45.8 | 39.7 | 73.2 | 50.5 | 31.1 | 36.7 |
| Fraction of days with ∂$^2$PM/∂VOC$^2$<0 | 0.42 | 41.4 | 3.3 | 62.1 | 3.8 | 38.7 | 13.5 | 79.9 |
| Fraction of days with ∂$^2$PM/∂NO$_x$∂VOC<0 | 5.8 | 6.7 | 1.3 | 41.8 | 21.4 | 29.1 | 6.6 | 74.0 |
| Fraction of days with ∂$^2$PM/∂T∂VOC<0 | 0.8 | 29.2 | 4.0 | 67.3 | 1.0 | 12.1 | 17.0 | 59.9 |

Figure 1. Evaluations of first derivatives of PM with respect to VOC made with the empirical models formed by neural networks with numbers of neurons ($N_N$) ranging from 21 to 25 (axes of abscissas), versus the similar evaluations made with the empirical models with $N_N$ ranging from 26 to 30 (axes of ordinates). Each point corresponds to one day at the training sub-set of the data-set. Solid lines represent best linear fits of the relationships.

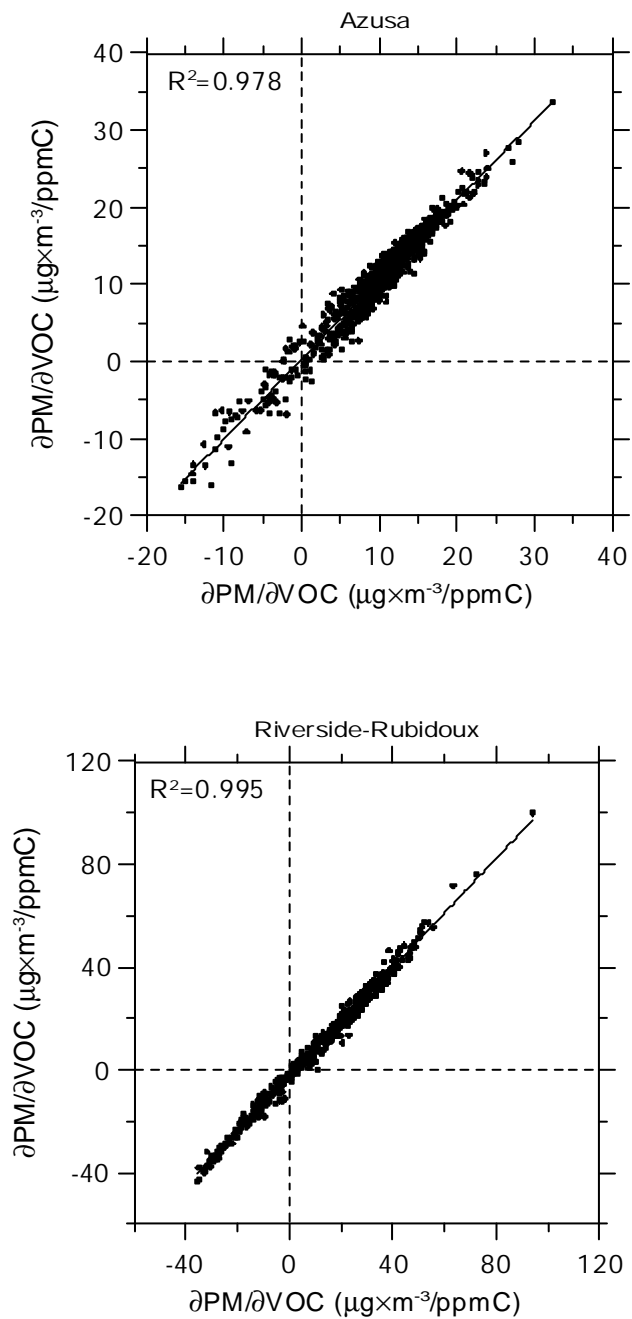

Figure 2. The empirical model evaluations of the PM concentrations (axes of ordinates) versus actually measured PM concentrations (axes of abscissas). Each point corresponds to one day at the training sub-set of the data-set. Solid lines represent best linear fits of the relationships.

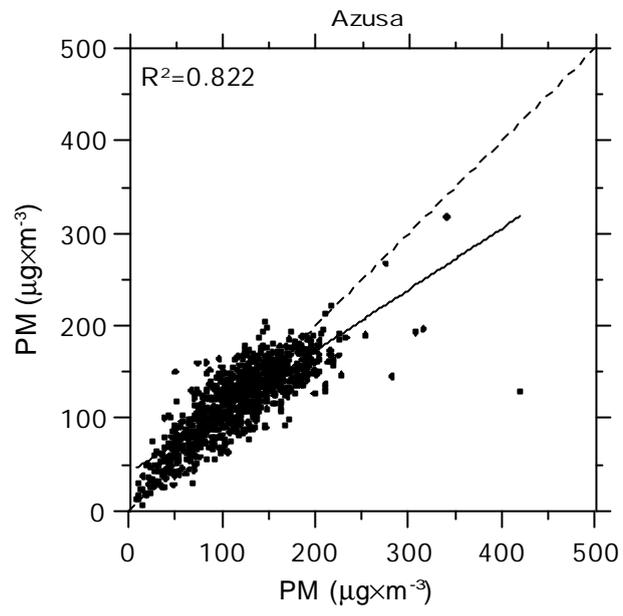

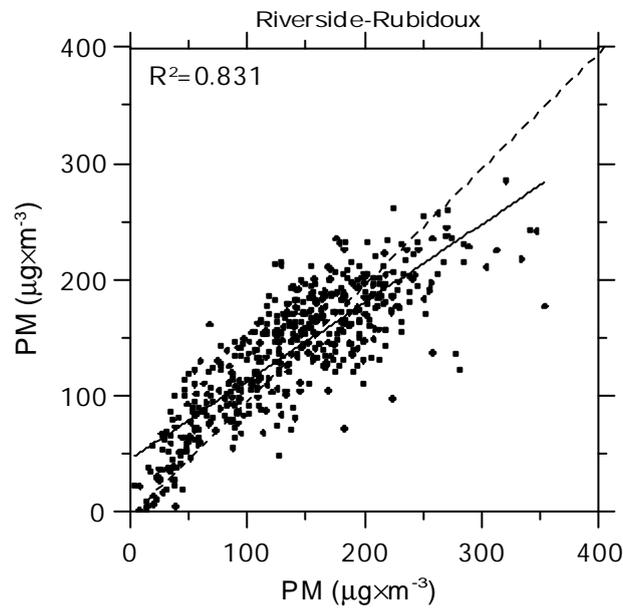

Figure 3. The dependencies of PM concentrations on $NO_x$ obtained from the empirical models for the days, for which negative sensitivity of PM to $NO_x$ is detected. $NO_x$ concentrations observed at these days are shown by crosses.

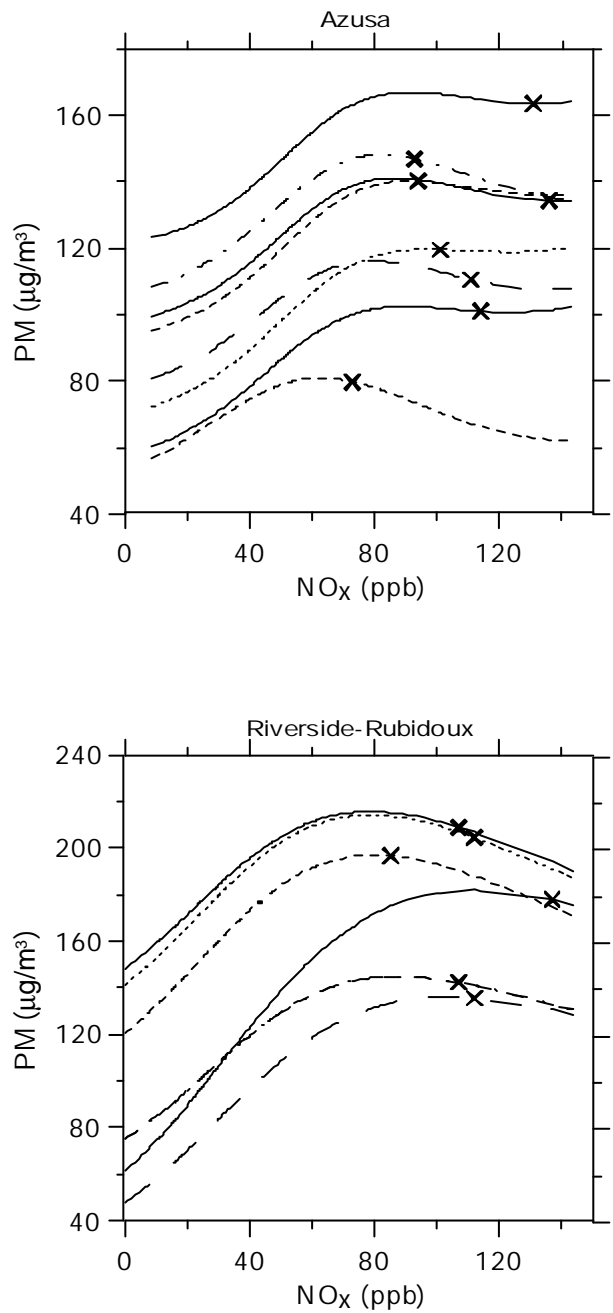

Figure 4. The empirical model evaluations of first derivatives of concentration of PM with respect to concentration of VOC, plotted against values of ambient temperature. The solid and dashed-dot lines represent 4th order polynomial fits of data and the boundary between two temperature modes, respectively. The dotted line marks the null level of the derivative values.

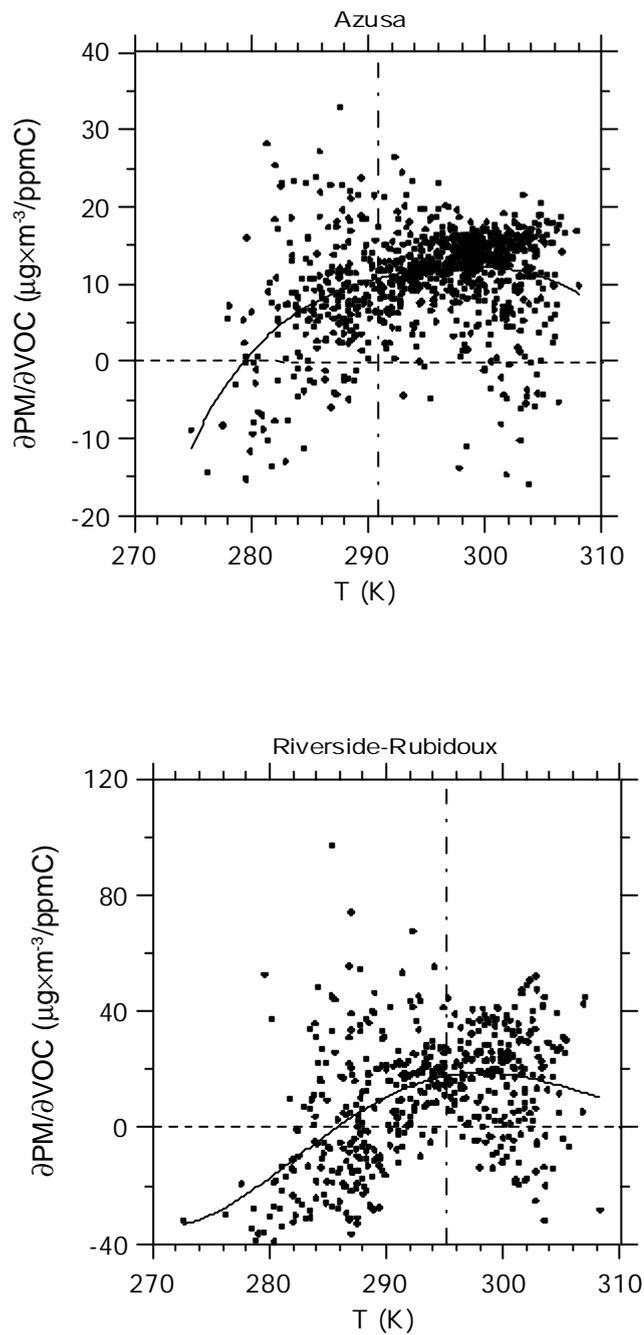

Figure 5. The empirical model evaluations of first derivatives of concentrations of PM with respect to concentration of VOC, plotted against values of ambient concentration of VOC. Dots and crosses correspond to low-temperature and high-temperature modes, respectively.

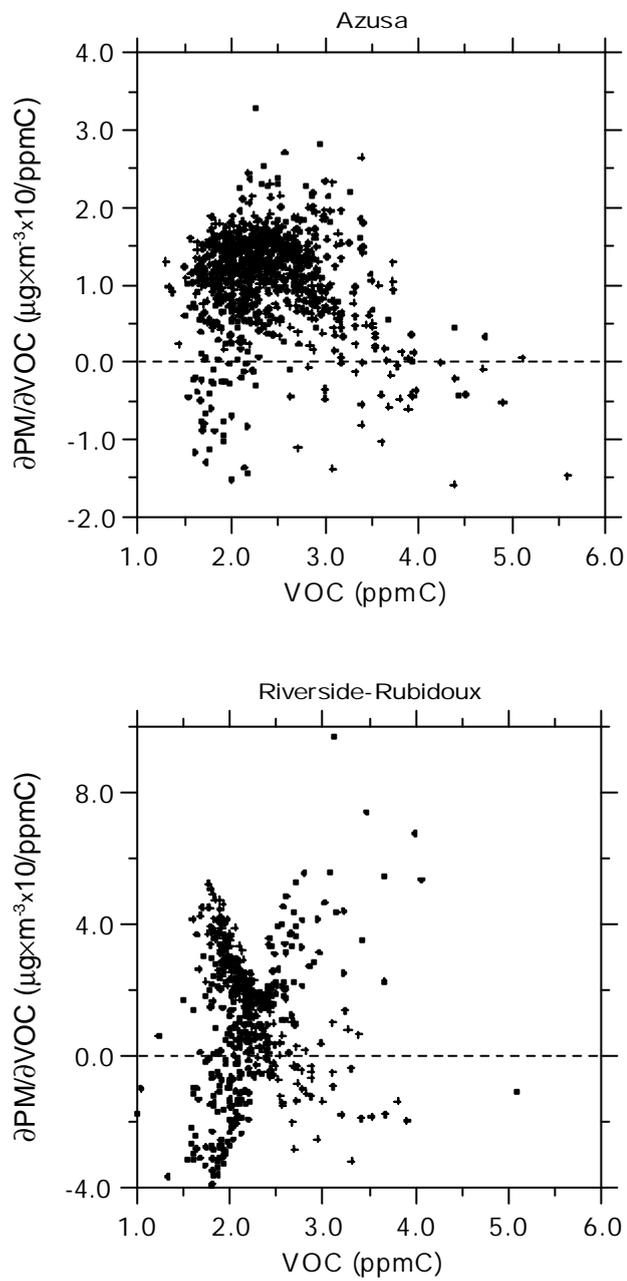